# The VIREO KIS at VBS 2018[*]

Final Notes[†]

Phuong Anh Nguyen [‡]
Yi-Jie Lu
Hao Zhang
Chong-Wah Ngo
VIREO Lab, Department of Computer Science,
City University of Hong Kong, HongKong SAR

## ABSTRACT
This short paper presents the video browsing tool of VIREO team which has been used in the Video Browser Showdown 2018. All added functions in the final version are introduced and experiences gained from the benchmark are also shared.

## 1 SYSTEM OVERVIEW

VIREO team started to join Video Browser Showdown [2] in 2017. The first version of the video browsing tool [3] is extended from SIRET's system [1] integrating with query-by-concept modal implemented for TRECVID MED zero-example event detection [4]. In 2018, the new version [5] was fully developed by including 3 main modalities for video searching which can be used independently or in combination to improve the search results:

- A compact *query-by-color-sketch* modal helps the user in sketching by providing the color distribution of video scenes in either frame or shot level with recommendation learned from the dataset.
- A *query-by-text* modal used for searching rich media surrounding and inside of videos including video description, speech and optical character recognized from the videos.
- A *query-by-concept* modal with an interface which allows the user directly select concepts from our large concept-bank. The user can also combine concepts by boolean operators AND, OR, NOT and assign weights easily using the designed interface.

An additional *relevance feedback* module is added providing incremental search based on positive shots selected by the user in the searching routine. Finally, the system included a browsing interface which compactly represents all candidate shots from each video as *a dynamic image* where the candidate shots are shown in sequence automatically on mouse hover (see Figure 1).

[*]This work was supported by two grants from the Research Grants Council of the Hong Kong Special Administrative Region, China (CityU 11210514, 11250716)
[†]The full version of paper is available at SpringerLink with doi:978-3-319-73600-6_42
[‡]Author's email: panguyen2-c@my.cityu.edu.hk

*MMM'18, February 2018, Bangkok, Thailand*
2018.

## 2 FINAL SYSTEM CHANGES

Online calculating all the possible combinations of color distribution given a user query is time consuming and not always necessary. The first change is to provide user an option to *turn on/off the recommendation module* to make it suitable for the benchmarking purpose which only lasts for 5 minutes per query.

Second, inspecting search result by mouse hover over videos to see dynamic images is not efficient for some queries. Hence, a flexible mechanism allows *switching browsing interfaces* between the proposed one and traditional video-frame-based interface is added.

A *query-by-object modal* with the same interface design as query-by-concept modal is added. This modal can detect 618 objects from video shots and was built on R-FCN using addition training data from multiples sources.

At last, we provide *filtering module* which can filter out black and white shots as well as black-bordered shots from the retrieved ranking list.

## 3 PERFORMANCE AT VBS 2018

VBS 2018 includes three types of tasks including visual known-item search, textual known-item search and ad-hoc video search. For known-item search task, the user needs to find a specific 20 seconds video segment in the corpus given a video segment (visual task) or a detail description (textual task). For ad-hoc video search task, the user tries to find as many correct video segments as possible given a general video description. VBS 2018 also hold two sessions for both expert and novice users to evaluate the effectiveness of the video browsing tool.

With the robustness of proposed query-by-color-sketch modal, we end up as the winner for visual known-item search task. As the final result announced, VIREO team stays at the 6$^{th}$ position over 9 team participated.

Looking back at the visual known-item search tasks, the users were able to complete a color-sketch query within 0.5 second. This allows user to input a combination of color distribution which leads to a small amount of results for inspecting.

In the textual known-item search tasks, we missed the correct shots for the given queries despite that these shots are ranked high. The log shows that the correct shots were returned in the top of our ranking list. The reason of missing is that the tool uses the master-shot key-frames provided by TRECVID for browsing, which overpasses the details relevant to query.

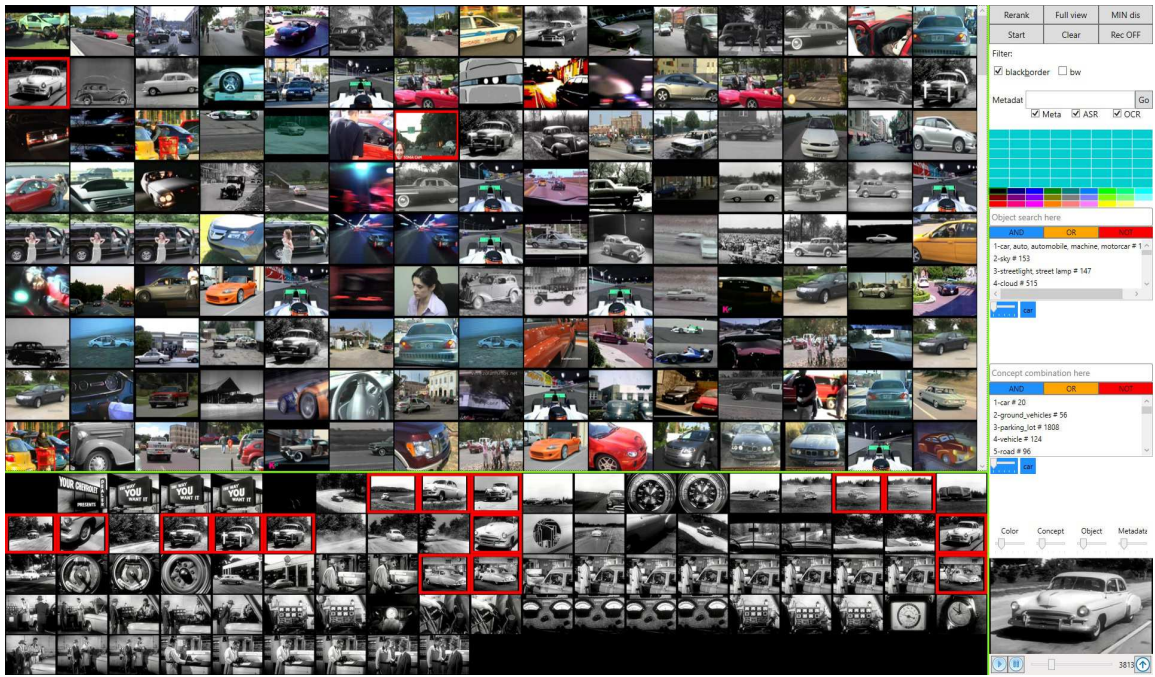

Figure 1: The VIREO's video browsing tool used in VBS 2018

In ad-hoc video search tasks, VIREO ranked at 2nd bottom up although we use a large concept bank with flexible selection and combination mechanism. There are two main reasons. First, the concept bank is large with different levels of meaning causing ambiguities in concept selection as well as unnecessarily back-and-forth result inspection and query reformulation. Seconds, the scoring system of VBS does not take into account the number of users participating in the benchmark, which disadvantages VIREO team who has only one user.

As observed in novice runs, the designed interface for formulating object and concept queries is not intuitive to beginner. User also can not modify query with ease. The browsing interface does not provide visual analytics helping user to understand search result, making user exhausted in hunting answers. Lastly, integrating multiple modalities may be helpful in some cases but also causes uncertainty for user which is the right modality to be selected for start searching.

## 4 CONCLUSIONS

As shown in our results, we conclude that our color-sketch modal works well for visual KIS task. Meanwhile, the interface for concept and object queries need to be improved avoiding confusing in query selection and combination. Also, the visualization techniques need to be developed and integrated into the system to enhance the browsing efficiency. These are interesting research direction and development to be considered in the coming VBS 2019.